%% file: bare_adv.tex
\documentclass[10pt,journal,compsoc]{IEEEtran}

\usepackage[boxed]{algorithm2e}
\usepackage{algpseudocode}
\usepackage{fullpage}
\usepackage{verbatim}
\usepackage{amsmath}
 \usepackage{booktabs}
\usepackage{subfigure}
\usepackage{graphics}
\usepackage[dvips]{graphicx}
\graphicspath{ {images/}}
\usepackage{hyperref}
\usepackage{caption}

\ifCLASSOPTIONcompsoc
  \usepackage[nocompress]{cite}
\else
  \usepackage{cite}
\fi
\ifCLASSINFOpdf
\else
\fi
\begin{document}
%
\title{A Machine Learning Approach to Quantitative Prosopography}
%
%
%
%

\author{Aayushee~Gupta, Haimonti~Dutta, Srikanta~Bedathur, Lipika~Dey}
\IEEEtitleabstractindextext{%

\begin{abstract}

Prosopography is an investigation of the common characteristics of a group of people in history, by a collective study of their lives. It involves a study of biographies to solve historical problems. If such biographies are unavailable, surviving documents and secondary biographical data are used. \textit{Quantitative} prosopography involves analysis of information from a wide variety of sources about ``ordinary people". In this paper, we present a machine learning framework for automatically designing a \textit{people gazetteer} which forms the basis of quantitative prosopographical research. The gazetteer is learnt from the noisy text of newspapers using a Named Entity Recognizer (NER). It is capable of identifying \emph{influential} people from it by making use of a custom designed Influential Person Index (IPI). Our corpus comprises of 14020 articles
from a local newspaper, ``The Sun", published from New York in 1896. Some influential people identified by our algorithm include Captain Donald Hankey (an English soldier), Dame Nellie Melba (an Australian operatic soprano), Hugh Allan (a Canadian shipping magnate) and Sir Hugh John McDonald (the first Prime Minister of Canada).

\end{abstract}

\begin{IEEEkeywords}
 Gazetteer, Text Mining, Information Retrieval, OCR, Spelling Correction, Historical data, Influential people detection.
\end{IEEEkeywords}}

\maketitle

\IEEEdisplaynontitleabstractindextext

%
\IEEEpeerreviewmaketitle

\ifCLASSOPTIONcompsoc
\IEEEraisesectionheading{\section{Introduction}\label{sec:introduction}}
\else
\section{Introduction}
\label{sec:introduction}
\fi

%
%
%
%
\IEEEPARstart{H}{istorical} newspaper archives provide a wealth of information. They are of particular interest to  historians \cite{allen2010historians}, genealogists (e.g. Genealogy Bank\footnote{\url{http://www.genealogybank.com/gbnk/}}, Ancestry\footnote{\url{http://www.ancestry.com/}}) and scholars.
An important use of historical newspapers is for People Search\cite{BilenkoMCRF03,Friedman_92} -- the process of finding information about a person and reconnecting them with others they are likely to know. The goal is to determine who knows whom and how. This is often achieved by studying biography. In historical groups, however, biographies may be largely untraceable. In such cases, secondary biographical information is studied by examination of the individual's experience and personal testimonies, some of which may be reported in newspaper articles. Identification of this group of individuals and studying the stories of their life is an important tool in the research historian's arsenal - called \textit{prosopography}. 
It can be used to learn social structure such as analysis of the roles of a certain group of people, holders of titles, members of professional and occupational groups or economic classes.
Quoting prosopographer Katharine Keats-Rohan, 
\begin{quote}
...prosopography is about what the analysis of the sum of data about many individuals can tell us about the different types of connection between them, and hence about how they operated within and upon the institutions -- social, political, legal, economic, intellectual -- of their time.
\end{quote}

The nature of prosopographical research has evolved over time. Lawrence Stone\cite{stone_71} discusses an ``older" form of prosopography which was principally concerned with well-known social elites, many of whom were influential people. Their genealogies were well-researched, and social webs and kinship linking could be traced, allowing a prosopography of a ``power elite" to emerge. This older prosopography can be contrasted with a newer form called \emph{quantitative prosopography}, which studied much wider populations including ``ordinary people". 

In this paper, we present a framework to develop a \textit{people gazetteer} which forms the basis of prosopographical research. The gazetteer is built from the text of historical newspapers subjected to Optical Character Recognition (OCR) and is capable of identifying influential people. Our paper has the following novel contributions: (1) \textbf{Development of the People Gazetteer} -- an organized dictionary of people names and a list of newspaper articles in which the name occurs. (2) \textbf{Identification of Influential People}: we define an Influential Person Index (IPI) which helps in identification and ranking of influential people.

To the best of our knowledge, the development of a framework for doing prosopographical research using machine learning has not been studied before. This exercise, however, opens up a wide range of possibilities -- for example, 
to build influential people networks that can learn about entities involved in historical events. Such applications can immensely help historians working on prosopography\cite{allen2013toward} and scholars in learning events related to historically significant people interactively.

\noindent \textbf{Paper Organization:} This paper is organized as follows: Section~\ref{influential:rw} discusses related work; the machine learning framework is discussed in Section~\ref{chp:framework}; the characteristics of the data used for this research are presented in Section~\ref{data}. Sections~\ref{chapter:people gazetteer} and ~\ref{influential} present the development of the gazetteer and the influential people detection process; empirical results and discussions are presented in Section~\ref{influential:results} and ~\ref{influential:discussion} and Section~\ref{conc} concludes the paper.

%

\renewcommand{\thetable}{\Roman{table}}
\renewcommand{\thefigure}{\arabic{figure}}

\section{Related Work}
\label{influential:rw}
\input{RelatedWork}

\section{Machine Learning Framework for Prosopographical Research}
\label{chp:framework}
\input{Framework}

\section{Dataset Description}
\label{data}
\input{Dataset}

\section{People Gazetteer}
\label{chapter:people gazetteer}
\input{PeopleGazetteer}

\section{Influential People Detection}
\label{influential}
\input{IPDetection}

\section{Results}
\label{influential:results}
\input{Results}

\section{Discussion}
\label{influential:discussion}
\input{Discussion}

\section{Conclusion}
\label{conc}
\input{Conclusion}
\bibliographystyle{IEEEtran}
\bibliography{aayushee}



\begin{IEEEbiography}[{\includegraphics[width=1in,height=1.25in,clip,keepaspectratio]{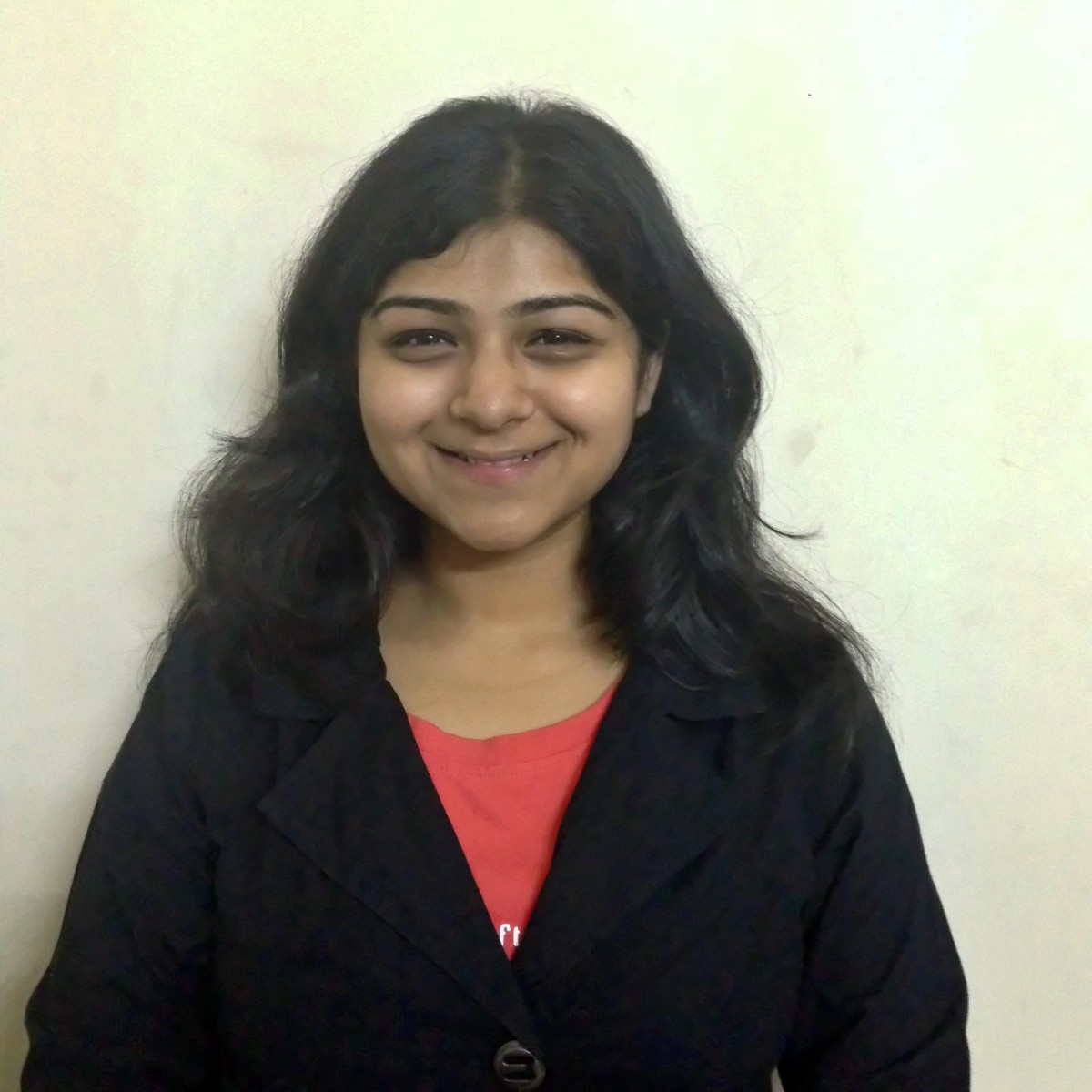}}]{Aayushee Gupta}
is an Assistant Professor in the Department of Computer Science and Information Technology at Jaypee Institute of Information Technology, India. She completed her M.Tech in Computer Science with specialization in Data Engineering from IIIT-Delhi. Her research interests include data mining and machine learning with applications in databases and information retrieval.
\end{IEEEbiography}


\begin{IEEEbiography}[{\includegraphics[width=1.5in,height=1.25in,clip,keepaspectratio]{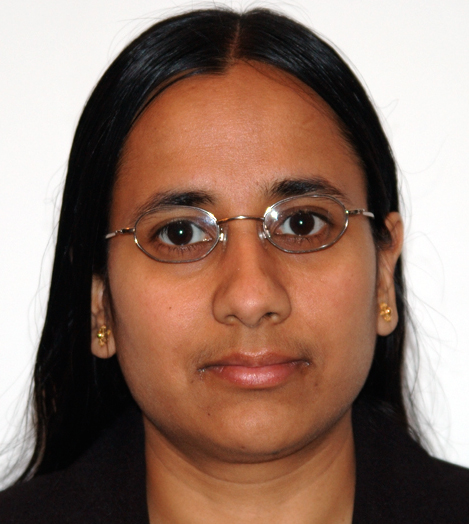}}]{Haimonti Dutta}
is an Assistant Professor at the Department of Management Science and Systems, University at Buffalo,USA. Prior to her current role, she was an Associate Research Scientist at the Center for Computational Learning Systems, Columbia University, NY where she headed the Scalable Analytics Research Group. Her research focuses on mining big data, machine learning and distributed optimization. 
\end{IEEEbiography}

\begin{IEEEbiography}[{\includegraphics[width=1in,height=1.25in,clip,keepaspectratio]{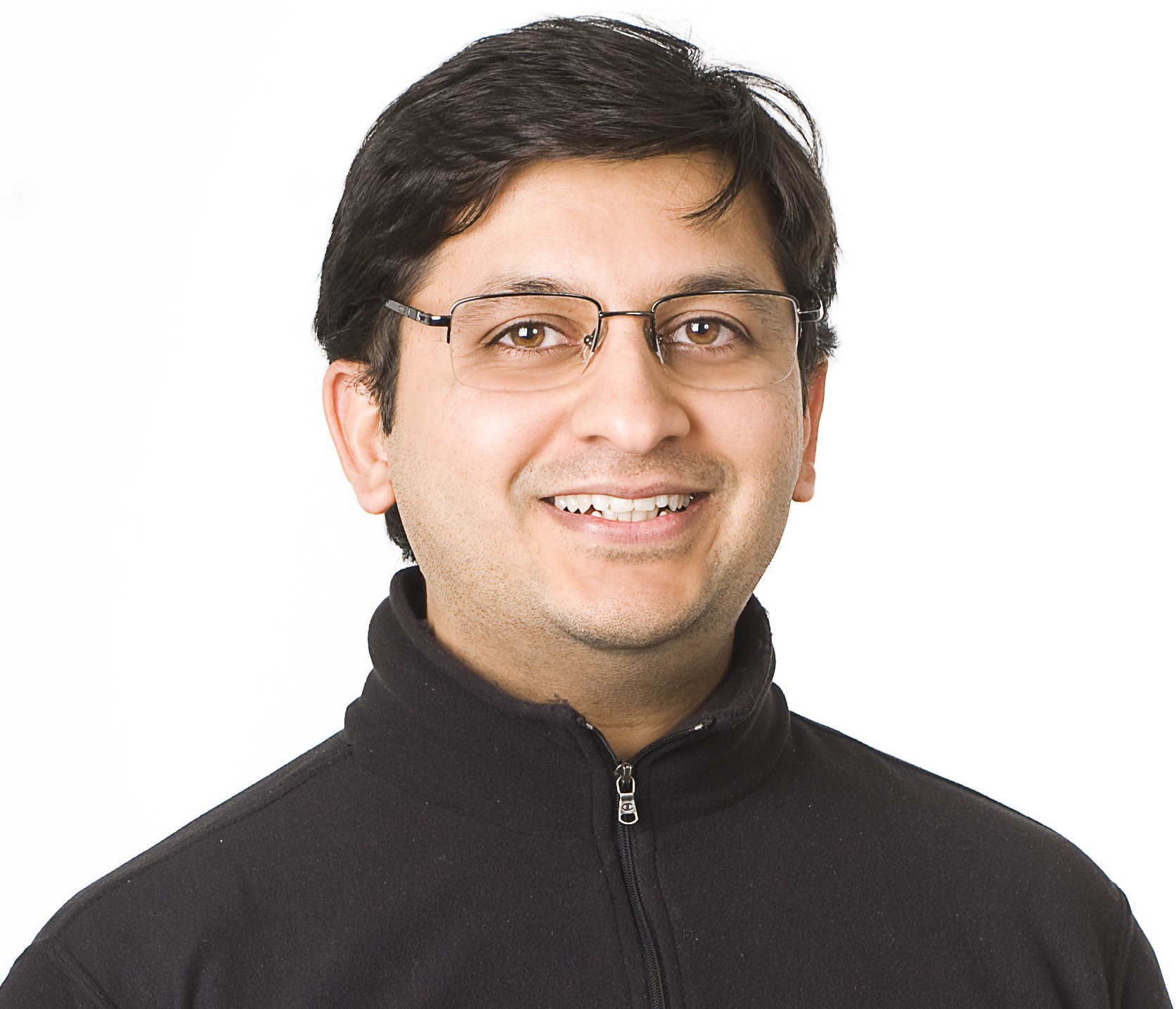}}]{Srikanta Bedathur}
received the Ph.D. degree from the Indian Institute of Science in Bangalore, India in 2005. He is currently a researcher at IBM Research India, and before that he was an Assistant Professor at IIIT-Delhi, where he was leading the Max-Planck 
Partner group on large-scale graph search and mining. He also held positions at Max-Planck Institute for Informatics and Saarland University in Germany. His current research primarily revolves around problems of scalable graph management and mining arising in large-scale knowledge repositories.
\end{IEEEbiography}

\begin{IEEEbiography}[{\includegraphics[width=1in,height=1.25in,clip,keepaspectratio]{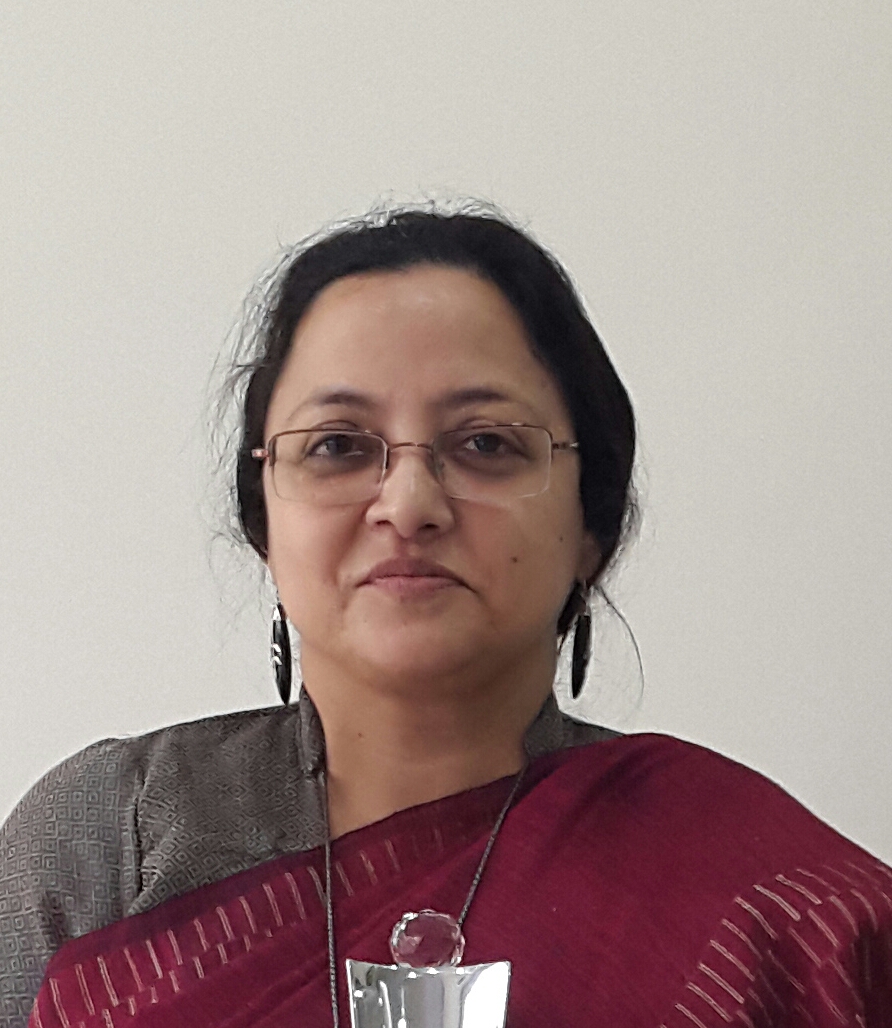}}]{Lipika Dey}
is a Senior Consultant and Principal Scientist at Tata Consultancy Services, India where she heads the Web Intelligence and Text Mining research group. Lipika did her Integrated M.Sc. in Mathematics, M.Tech in Computer Science and Data Processing and Ph.D. in Computer Science and Engineering - all from IIT Kharagpur. Prior to joining TCS in 2007, she was a faculty at the Department of Mathematics at IIT Delhi from 1995 to 2007. Lipika's research interests are in the areas of content analytics from social media and other consumer-generated text, social network analytics, predictive modeling, sentiment analysis and opinion mining, and semantic search of enterprise content. Her focus is on seamless integration of social intelligence and business intelligence using multi-structured data analytics.
\end{IEEEbiography}





\begin{table*}
\caption{Table showing Topic ID and words obtained from the 30 Topics LDA model.}
\label{table:topicwords}
\resizebox{16cm}{!} {
 
   \begin{tabular}{|p{1cm}|p{16cm}|}

    \hline
    TOPIC ID & TOPIC WORDS                                                                                                                                           \\ \hline
    1        & total ii won club score night ran furlough alleys tournament time   mile fourth rolled curling scores race national game                              \\ \hline
    2        & la lu ot lo tu au tb ta ha tea day al aa ut ar uu wa tt te                                                                                            \\ \hline
    3        & iii lie tin nail tn lit hut ill ii nn thu tu anti thin inn hit lu lo   nut                                                                            \\ \hline
    4        & line street feet point western easterly northerly feel southerly   distance place distant lo fret hue beginning laid early felt                       \\ \hline
    5        & opera theatre music company week play stage evening night performance   concert mme audience manager season de orchestra house miss                   \\ \hline
    6        & great people life man women good country world american part ot ha   made la years make long place bad                                                \\ \hline
    7        & election mr party republican state district vote democratic county senator   elected city committee mayor political candidate majority york democrats \\ \hline
    8        & time ho work tn men city bo lie anti day thin long thu made part ago   lot york make                                                                  \\ \hline
    9        & st room av sun wife board front lo december rent lot november sunday   ht west ar house private si                                                    \\ \hline
    10       & dr book st story books cloth author cure free work york blood   illustrated remedy goods medical library health price                                 \\ \hline
    11       & church dr father funeral school st college sunday year rev catholic pastor services late service held society holy clock                              \\ \hline
    12       & horse race class horses won racing years prize record year show ring track mile money jockey trotting trotter ran                                     \\ \hline
    13       & cent year week pf market total net stock today central st ft lit sales short cotton ohio lot month                                                    \\ \hline
    14       & white water indian black long found thu big dog time ground wild tree killed birds bird day great lake                                                \\ \hline
    15       & price black silk goods prices ladies worth dress fine white full tea quality style wool made fancy cloth fur                                          \\ \hline
    16       & street mrs mr avenue wife house miss yesterday years home woman night ago husband found died daughter children mother                                 \\ \hline
    17       & war american government army chinese japanese china japan foreign united nov emperor states prince minister military french port navy                 \\ \hline
    18       & feet north minutes avenue boundary seconds degrees west york minute degree point east south feel city angle county laid                               \\ \hline
    19       & man ho men night back wa room left house told bad door found turned place ran lie front morning                                                       \\ \hline
    20       & water feet building boat company car train road fire miles railroad island work line city great river built bridge                                    \\ \hline
    21       & club game team play football half ball left college back yale played harvard line eleven men match yacht field                                        \\ \hline
    22       & ii iii ill lit ll si ti il im vi st iv ft mi li till lull lui oil                                                                                     \\ \hline
    23       & bank money national gold amount notes banks hank business treasury account cent paid bonds note currency company stock estate                         \\ \hline
    24       & mr john william york henry charles james club city ii george dec dr thomas smith jr brooklyn van held                                                 \\ \hline
    25       & piano st rooms car york daily chicago city sunday upright parlor furnished broadway hotel av west train brooklyn monthly                              \\ \hline
    26       & york daily steamship nov directed letter dec fur orleans al steamer walls letters close australia china japan city london                             \\ \hline
    27       & mr court police judge justice case yesterday street district witness jury charge asked attorney trial arrested lawyer told office                     \\ \hline
    28       & mr law present made public year state committee president secretary bill report states con tin united number meeting york                             \\ \hline
    29       & air ran ur fur ui full tt al tl late mr ant liar art lay told met ti tr                                                                               \\ \hline
    30       & company york trust bonds city cent railroad mortgage interest wall bond stock street st central january coupon committee jan                          \\ \hline
    \end{tabular}}

\end{table*}

\begin{table*}
\caption{Table representing top 30 influential person entities detected from people gazetteer with 30 Topics LDA Model along with evaluation results and comments.}
\label{table:app1}
\centering
\begin{tabular}{|l|l|p{3cm}|p{3cm}|}
\hline
\textbf{Person Name}      & \textbf{IPI}      & \textbf{Whether found on Wikipedia} & \textbf{Comments}                                     \\ \hline
capt creeten	 & 3.33 & no                 & spelled incorrectly;capt creedon \\ \hline
capt hankey      & 3.05 & yes                &                                      \\ \hline
capt pinckney    & 2.93 & yes                &                                      \\ \hline
john martin   & 2.89 & yes                &                                      \\ \hline
ann arbor	& 2.87	& no & location name		\\ \hline
john macdonald      & 2.85 & yes                &                                      \\ \hline
aaron trow       & 2.81 & yes                & fictional character                  \\ \hline
mrs oakes        & 2.79 & no                 & false positive                       \\ \hline
alexander iii    & 2.71 & yes                &                                      \\ \hline
buenos ayres     & 2.70 & no                 & location name                            \\ \hline
mrs martin       & 2.70 & no                 & false positive                       \\ \hline
caleb morton     & 2.63  & no                 & fictional character                  \\ \hline
anthony comstock & 2.63 & yes                &                                      \\ \hline
john thompson    & 2.61 & yes                &                                      \\ \hline
nat lead         & 2.61 & no                 & false positive                       \\ \hline
van cortlandt    & 2.54 & no                 & location                             \\ \hline
ed kearney       & 2.54 & yes                & name of horse                        \\ \hline
louis philippe   & 2.52 & yes                &                                      \\ \hline
mrs talboys      & 2.52 & yes                & fictional character                  \\ \hline
jim hooker       & 2.50 & yes                & false positive                       \\ \hline
marie clavero    & 2.49 & no                 & false positive                       \\ \hline
charley grant    & 2.45 & no                 &                     \\ \hline
james mccutcheon & 2.43 & no                 & part of an organization name                      \\ \hline
hugh allan       & 2.43   & yes                &                                      \\ \hline
william i        & 2.42   & yes                &                                      \\ \hline
marie antoinette & 2.41  & yes                &                                      \\ \hline
mr john      & 2.39 & no              & false positive                       \\ \hline
schmitt berger   & 2.39 & no                 & spelled incorrectly;max f schmittberger                       \\ \hline
jacob schaefer   & 2.39 & yes                &                                      \\ \hline
phil king	&	2.38	& yes	& 				\\ \hline
             
\end{tabular}

\end{table*}

\begin{table*}
\caption{Table representing top 30 influential person entities detected from people gazetteer with 100 Topics LDA Model along with evaluation results and comments.}
\label{table:app2}
\centering
\begin{tabular}{|l|l|p{3cm}|p{3cm}|}
\hline
\textbf{Person Name}      & \textbf{IPI}      & \textbf{Whether found on Wikipedia} & \textbf{Comments}                                     \\ \hline
capt creeten & 3.28 & no & spelled incorrectly; capt creedon \\ \hline
mrs martin & 3.21 & no & false positive \\ \hline
capt hankey & 3.19 & yes &  \\ \hline
alexander iii & 3.05 & yes &  \\ \hline
aaron trow & 2.79 & yes &  \\ \hline
john martin & 2.78 & yes &  \\ \hline
john macdonald & 2.77 & no &  \\ \hline
mrs oakes & 2.74 & no & false positive \\ \hline
ed kearney & 2.63 & yes & name of horse \\ \hline
caleb morton & 2.61 & no & fictional character \\ \hline
nat lead & 2.58 & no & false positive \\ \hline
john ward & 2.57 & yes &  \\ \hline
van cortlandt & 2.50 & no & location \\ \hline
mrs talboys & 2.49 & yes & fictional character \\ \hline
ann arbor & 2.49 & no & location \\ \hline
buenos ayres & 2.49 & no & location \\ \hline
john thompson & 2.48 & yes &  \\ \hline
louis philippe & 2.47 & yes &  \\ \hline
marie clavero & 2.47 & no & false positive \\ \hline
hardy fox & 2.45 & no &  \\ \hline
charley grant & 2.42 & no & \\ \hline
mme melba & 2.41 & yes &  \\ \hline
charles weisman & 2.40 & no & false positive \\ \hline
hugh allan & 2.40 & yes &  \\ \hline
henry w dreyer & 2.39 & no &  \\ \hline
schmitt berger & 2.37 & no & spelled incorrectly \\ \hline
phil king & 2.36 & yes &  \\ \hline
henry a meyer & 2.35 & no &  \\ \hline
north orlich & 2.35 & no & false positive \\ \hline
james mccutcheon & 2.34 & no & part of organization name \\ \hline
\end{tabular}

\end{table*}

\end{document}

%% file: RelatedWork.tex
In this section, we review two types of related literature - digital humanities projects which build gazetteers from text and the process of identification of influential people from data.

\subsection{Gazetteers for Digital Humanities Projects}
Newspaper archives have been studied extensively for the design of search and retrieval algorithms (\cite{Shahaf_11, Gabrilovich_04a, Alonso_10, Khurdiya_11}), summarization(\cite{McKeown95, Otterbacher06, Radev97, Radev01, Radev05}), sentiment analysis (\cite{balahur2009rethinking, godbole2007large, li2014sentiment}), topic modeling (\cite{Masand_92, Nallapati_04a, Radev99c, au2011studying, lee2010topic}), clustering(\cite{dutta2011learning}), classification (\cite{dutta2012using} and visualization(\cite{torget2011mapping},\cite{southall2014pastplace}). The National Digital Newspaper Program (NDNP) in the United States\footnote{This is a partnership between the National Endowment for the Humanities (NEH) and the Library of Congress (LC).}, is a long-term effort to develop a searchable database of U.S. newspapers. Historical newspapers available from this project (named Chronicling America\footnote{http://chroniclingamerica.loc.gov/}), have been used for topic modeling\cite{yang2011topic}. Newman et. al\cite{newman2006analyzing} use a combination of Statistical Topic Modeling and Named Entity Recognition for analyzing entities and topics. 
Lloyd et. al \cite{lloyd2005lydia} discuss their approach for designing a news analysis system\footnote{http://www.textmap.com} where information about several types of entities can be searched. They perform temporal and spatial analysis and present time series popularity graphs based on the number of reference and co-reference names for the entity. 

Several digital humanities projects that have used machine learning and natural language processing techniques to learn from historic newspaper archives are relevant to this work -- the libraries of Richmond and Tufts have examined the Richmond Times Dispatch during the civil war years for more than two decades and their work focuses on automatic identification and analysis of full OCR text in newspapers to provide advanced searching, browsing and visualization\cite{crane2006challenge,smith2002detectinga, smith2002detectingb, smith2001disambiguating}. 

Developing gazetteers from news articles is a well established technique - different types of gazetteers are discussed under the General Architecture for Text Engineering (GATE\footnote{http://gate.ac.uk/sale/tao/splitch13.html}) framework. It defines a gazetteer as a set of lists containing names of entities (such as cities, organizations, days of the week, etc) which can be used to find occurrences in the text. We use this definition to develop our People Gazetteer that finds person name entities from a news article repository and associates each unique person entity with the list of articles in which they occur.

Gazetteer lists are also discussed in \cite{carlson2009learning} where they are used for learning named entity taggers using partial perceptron and aid in performing better NER compared to CRF based entity taggers. Zhang et. al\cite{zhang2009novel} discuss automatic generation of gazetteer lists by finding entities with similar labels from Wikipedia articles. 
Allen et. al \cite{allen2013toward} describe an exploratory study for developing an interactive directory for the town of Norfolk, Nebraska for the years 1899 and 1900. Their work focuses on providing structured and richer information about the person entities by linking their occurrences with associated events described in historical newspapers. 

\subsection{Influential People Detection}
In prosopography, identification of the ``social elite" plays an important role. Their experience and personal testimonies may be reported at length in newspaper articles. 

In the context of machine learning, influential people detection has been mostly done in the field of social networks, marketing and diffusion research.
Kempe et. al \cite{kempe2003maximizing} present work on choosing the most influential set of nodes in a social network in order to maximize user influence in the network. 
Lerman et. al \cite{lerman2010using} define popularity of a news story in terms of number of reader votes received by it. Popularity over time is based on voting history and the probability that a user in a list will vote. To identify influential bloggers, Agarwal et. al\cite{agarwal2008identifying} quantify influence of each blogger by taking the maximum of the influence scores of each blog posted by the blogger. The influence score is calculated using the number of posts that refer to the blog, number of comments on the blog, number of other posts that the blog refers to and length of the blog. Influential blogger categories are also created based on the temporal patterns of blog posting. Cha et. al\cite{cha2010measuring} describe another set of measures for detection of top influential users on Twitter using number of retweets, mentions and followers for an individual. They perform ranking based on each measure separately and use Spearman's rank correlation coefficient to find correlation among ranks and effect of each measure contributing to a person's influence. The influence ranks of topmost influential users on Twitter are presented across various topics as well as time. In all of the above, the goal is to measure influence or popularity -- however, these cannot be directly adapted to the gazetteer or newspaper articles. 

%% file: Framework.tex
\begin{figure}
\centering
\includegraphics[width=0.5\textwidth, height=0.6\textheight]{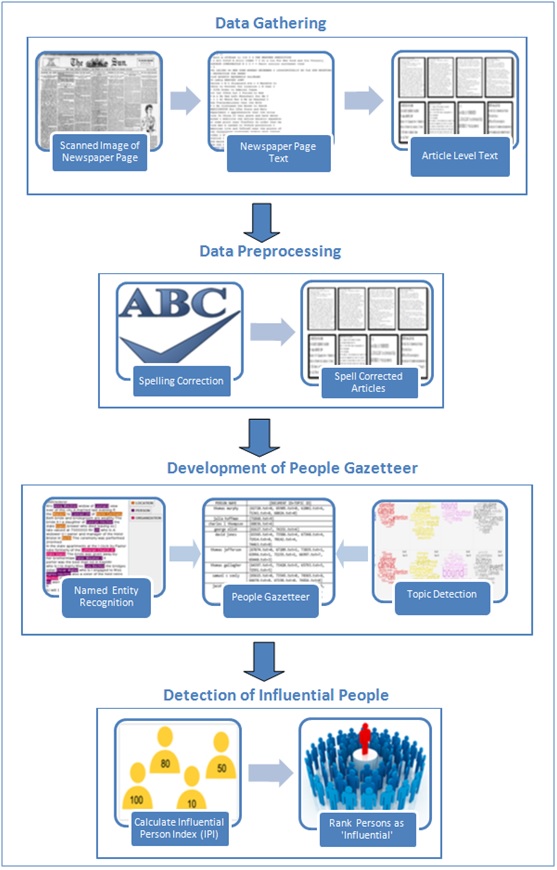}
\caption{Research Framework showing components of proposed solution}
\label{fig:framework}
\vspace{-10pt}
\end{figure}
Figure~\ref{fig:framework} presents the framework for machine learning to aid prosopographical research. It has the following components:
(1)  \textbf{Data Gathering: } Prosopographical studies involve research on biographies of a group of people and is therefore severely limited by the quantity and quality of data accumulated about the past. Often in historical groups, a lot of information is available about some people, and almost nothing about some others. Studies are severely affected by lack of information and hence secondary sources of information are resorted to including demographic sources (such as parish registers), economic sources (such as deeds of sales), fiscal sources (such as tax lists), financial sources (such as city accounts), administrative sources (such as company records), religious sources (such as membership lists of fraternities), judicial sources (such as sentences), family archives and photographs, publicly available information (such as newspaper archives). The context of the research is sketched based on the available literature and has to be sufficient, relevant and easily accessible. 
For this research, the primary source are digitized newspaper archives. In order to make a newspaper available for searching on the Internet,
the following processes \cite{dutta2011learning} must take place: (1) the microfilm copy of original paper is scanned; (2) master and Web image files are
generated; (3) metadata is assigned for each page to improve the
search capability of the newspaper; (4) OCR software is run over high
resolution images to create searchable full text and (5) OCR text,
images, and metadata are imported into a digital library software
program. 
(2) \textbf{Data Pre-processing: } The images obtained from the OCR software are manually segmented to obtain article level data. 
Following this, several preprocessing steps are applied on the text of the news
articles including  spelling correction and evaluation using a novel algorithm presented in \cite{Gupta_14a}.
(3)  \textbf{Development of the People Gazetteer: }This component describes the process of development
of the people gazetteer which involves Named Entity Recognition (NER) in order to find person
entities. This is followed by topic detection using Latent Dirichlet Allocation (LDA) to find the primary topic(s) of news articles
and both are linked to obtain an organized structure.
(4) \textbf{Detection of Influential People: } This component defines an \textbf{Influential Person Index}
(IPI) that incorporates several criteria for identifying and ranking of influential
people. Details about IPI, ranking and final results with some case studies are discussed in Section~\ref{influential:results}.

%% file: Dataset.tex
Our prosopographical research is based on historical newspapers obtained from Chronicling America\footnote{\texttt{http://chroniclingamerica.loc.gov/}}. 
Under this program, institutions such as libraries receive an award to select and digitize approximately 100,000 newspaper pages representing that state's regional history, geographic coverage, and events of the particular time period being covered. The scanned newspaper holdings of the New York Public Library are a source of prosopographical studies. 

\begin{figure*}
\includegraphics[scale=0.75]{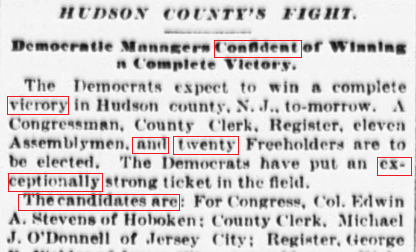}
\includegraphics[scale=0.80]{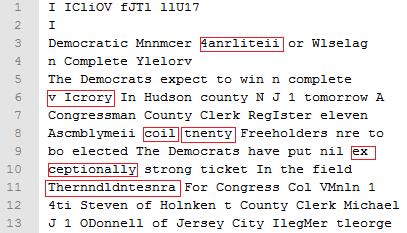}
\caption{Scanned Image of a Newspaper article (left) and its OCR raw text (right)}
\vspace{-10pt}
\label{figure:1}

\end{figure*}

\subsection{Characteristics}
The newspapers are scanned on a page-by-page basis and article level
segmentation is poor or non-existent; the OCR scanning process is far
from perfect and the documents generated from it contain a large
amount of garbled text.
An individual OCR text article has at least one or more of the following types of spelling errors:
 \textbf{1) Real word errors} include words that are spelled correctly in the OCR text but still incorrect when compared to the original newspaper article image.  \textbf{2) Non-real word errors} include words that have been misspelled due to some insertion, deletion, substitution or transposition of characters from a word.  \textbf{3) Non-word errors} include words that have been spelled incorrectly and are a combination of alphabets and numerical characters. \textbf{4) New Line errors} include words that are separated by hyphens where part of a word is written on one text line and remaining part in the next line. 
\textbf{5) Word Split and Join errors} include words that either get split into one or more parts or some words in a sentence get joined to a make a single word. 
For example, in Figure~\ref{figure:1} the word ``coil"  has been correctly spelled in the OCR text but should have been ``and" according to the original newspaper article (real word error); the word ``tnenty" in the OCR text has a substitution error (`n' should have been `w') which is actually ``twenty" according to the original newspaper article (Non-real word error); the word ``4anrliteii" is a combination of alphabets and number and should have been ``confident" as per the original newspaper article(Non-word error); the word ``ex-ceptionally" where ``ex" occurs on one line while ``ceptionally" in the next and due to no punctuation in the text, they are treated as separate words in OCR text(New Line error); the word ``Thernndldntesnra" in the OCR text is actually a combination of three words ``The candidates are" while the words ``v Icrory" are actually equivalent to a single word ``victory" when compared with the original news article(Word Split and Join error).

\subsection{Statistics}
Article level segmentation of text is available for only two months -- since this requires human intervention. Articles of ``The Sun" newspaper from November-December 1894 consisting of 14020 news articles are used in our study. A total of 8,403,844 tokens are generated from a bag-of-words extraction. 
The text from the articles do not have any punctuation and contain a large amount of garbled text containing above mentioned OCR errors.

\subsection{Preprocessing}
The garbled OCR text makes data preprocessing mandatory before application of any text mining algorithms. We,therefore, use edit distance algorithm based on Levenshtein distance to perform spelling correction on the OCR text articles. The algorithm is chosen because of its speed and ability to correct OCR errors compared to the n-gram approach \cite{chattopadhyaya2013fast}. Our edit distance algorithm also uses an enhanced person names dictionary for look up to give significance to personal names spelling correction in the dataset. The results of spelling correction and data preprocessing are presented in \cite{Gupta_14a}.

%% file: PeopleGazetteer.tex
People Gazetteer as defined in Section~\ref{sec:introduction} consists of tuples of person names along with list of documents in which they occur. The primary goal of developing the gazetteer is to have an organized list of person names from which influential people can be identified.
This section describes the two-step process involved in the construction of the People Gazetteer:
a) Extraction of person names from the news articles using Named Entity Recognition (described further in Section~\ref{ner}) and
b) Assignment of topics to news articles using a Latent Dirichlet Allocation (LDA)-based topic detector (described further in  Section~\ref{topic detection}).

\subsection{Person Named Entity Recognition (PNER)}
\label{ner}

\noindent Named Entity Recognition (NER) refers to classification of elements in text into pre-defined categories such as the names of persons, organizations, locations, expressions of times, quantities, monetary values, and percentages. 
Person Named Entity Recognition (PNER) is Named Entity Recognition that marks up only person names occurring in the text. It involves \emph{chunking} or segmentation of  the text for name detection followed by \emph{classification} of the name by entity type ( for e.g. person, organization, and location).


In this work, the Stanford CRF-NER\footnote{http://nlp.stanford.edu/software/CRF-NER.shtml} is used for Person Named Entity Recognition. It identifies three classes -- Person, Organization and Location and is based on linear chain Conditional Random Field (CRF)(\cite{mccallum2003early, finkel2005incorporating, sutton2011introduction}) sequence models. 
The following modifications are made on the output of the Stanford CRF-NER -- if the software recognizes single term person entities, we ignore those and consider only multi-term person entities and if a person entity's name is repeated multiple times, then we consider it only once. For example, if the person names ``John", ``John Smith", ``Smith" are recognized during PNER process, then we only consider ``John Smith" as as a potential person entity for our People Gazetteer.  
A total of 36364 person entities are extracted from our corpus of 14020 news articles.  
The person name entities are binned into the following categories based on the number of news articles of their occurrence:
(1)\textbf{Not Influential}: Person entities with occurrence in less than 4 news articles. (36004 person entities )
(2)\textbf{Popular}: Person entities with occurrence from 4 to 15 news articles. (344 person entities) 
(3)\textbf{Elite} : Person entities with occurrence in 16 or more news articles. (16 person entities)
The categories defined above have been chosen manually and simply provide a mechanism for grouping people. It does not lead directly to the conclusion that a person entity with large number of articles is influential. Also, slight perturbations to these values (4 and 15) does not change the final results of ranked influential people.


\subsection{Topic Detection}
\label{topic detection}


Topic detection involves the process of identifying topics from a document collection using a topic model\cite{blei2012probabilistic}.  
The following algorithm is run over all the news articles in our repository.




\noindent \textbf{Latent Dirichlet Allocation (LDA)}
LDA is a generative probabilistic model in which each document is modeled as a finite mixture over an underlying set of topics and each topic, in turn, is modeled as an infinite mixture over an underlying set of topic probabilities\cite{blei2003latent}. 
Given an input corpus of $D$ documents with $K$ topics, each topic being a multinomial distribution over a vocabulary of  $W$ words, the documents are modeled by fitting parameters ${\Phi}$ and ${\Theta}$. ${\Phi}$ is a matrix of size $D \times K$ in which each row is a multinomial distribution of document $d$  indicating the relative importance of words in topics. ${\Theta}$ is the matrix of size $W \times K$ with each column a multinomial distribution of topic $j$ and corresponds to the relative importance of topics in documents. Given the observed words x = ${x_i}_j$, inference in LDA is done by computing the
posterior distribution over the latent topic assignments z = ${z_i}_j$, the mixing proportions ${\Theta_j}$  and the
topics ${\Phi_k}$.  The inferencing is either done using variational bayesian methods or Gibbs sampling which involves integration and sampling of latent variables.
However, LDA is a compute intensive algorithm and it can take several days to run over a large corpora.

\noindent \textbf{Distributed LDA Model: } In practice, to scale LDA a parallel algorithm is used. The data is partitioned among processors and inference is done in parallel. 
The Approximate Distributed LDA (AD--LDA) model (\cite{newman2009distributed}) assumes the dataset $D$ is distributed equally among $P$ processors. A random assignment of topics is made to each processor so that it has its own copy of words $x_p$, topics $z_p$, word topic counts ${{{N_w}_k}_p}$ and topic counts ${{{N_k}_j}_p}$. 
Gibbs sampling is used for inferencing local topic models on each processor for a given number of iterations and topic probabilities $z_p$, word topic ${{N_w}_k}_p$ and topic counts ${{N_k}_j}_p$ are reassigned.
Global update is performed after each pass by using a reduce-scatter operation on word topic count ${{N_w}_k}_p$ to get a single set of counts and obtain final topic assignments.

\noindent \textbf{How good are these topic models?}
Topic models can be evaluated using \emph{perplexity} (\cite{newman2009distributed, blei2003latent}) which expresses how surprised a trained model is, when given unseen test data. 
Formally, perplexity can be calculated using the following formula:

$$Perplexity= \exp(-\dfrac{\text{Log Likelihood of unseen test data}}{\text{Number of tokens in the test set}})$$
Perplexity is a decreasing function of the log likelihood of the unseen documents and lower the perplexity, better is the topic model.

\subsection{People Gazetteer Output }
\label{gaz:result}


\begin{figure}[h]
\centerline{\includegraphics[width=0.50\textwidth]{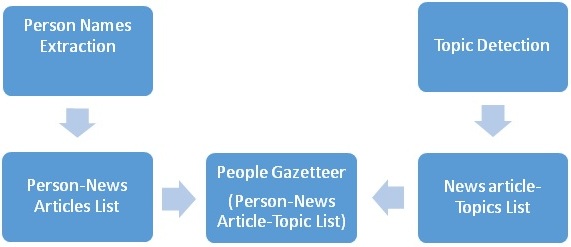}}
\caption{Procedure for development of People Gazetteer}
\label{figure:gaze}
\end{figure}

The procedure of development of the people gazetteer can be seen in Figure~\ref{figure:gaze}. The list of articles obtained for each person entity after application of PNER (Person-Article List) and the highest scoring topic assigned to it during topic detection (Article-Topic List) are combined to obtain an entry in the People Gazetteer. 
A snapshot is illustrated in Table~\ref{table:gaz}.

\begin{table*}

  \begin{minipage}[b]{\linewidth}
  \begin{center}
 \begin{tabular}{|p{2.5cm}|p{13cm}|}
  \hline
    PERSON ENTITY NAME & DOCUMENT LIST (Document ID $\rightarrow$  Document Topic ID)\\ \hline
    Thomas Murphy        & (61720.txt$\rightarrow$16, 62002.txt$\rightarrow$11, 65905.txt$\rightarrow$19, 71341.txt$\rightarrow$28, 68024.txt$\rightarrow$16)                        \\ \hline
    George Eliot       & (74151.txt$\rightarrow$5, 61627$\rightarrow$15)                                                                                      \\ \hline
    Charles L Thompson        & (68836.txt$\rightarrow$9)                                                                         \\ \hline
    Thomas Jefferson       & (67874.txt$\rightarrow$19, 67209.txt$\rightarrow$28, 63996.txt$\rightarrow$6, 73835.txt$\rightarrow$6, 71155.txt$\rightarrow$6, 65440.txt$\rightarrow$5, 66997.txt$\rightarrow$20)                   \\ \hline
    Jacob Schaefer        & (70205.txt$\rightarrow$21, 63936.txt$\rightarrow$22, 68554.txt$\rightarrow$21, 73420.txt$\rightarrow$21, 74550.txt$\rightarrow$21, 74922.txt$\rightarrow$21, 64577.txt$\rightarrow$21, 74759.txt$\rightarrow$21, 67340.txt$\rightarrow$0, 67924.txt$\rightarrow$2)                   \\ \hline
    Queen Victoria       & (68231.txt$\rightarrow$5, 74775.txt$\rightarrow$5, 75097.txt$\rightarrow$5, 72221.txt$\rightarrow$2, 62731.txt$\rightarrow$5, 62616.txt$\rightarrow$17, 68368.txt$\rightarrow$17)                                      \\ \hline
    Thomas Gallagher        & (64397.txt$\rightarrow$28, 65793.txt$\rightarrow$21, 72591.txt$\rightarrow$0, 73420.txt$\rightarrow$21) \\ \hline
    Samuel S Seely     & (70365.txt$\rightarrow$2, 64670.txt$\rightarrow$23, 65615.txt$\rightarrow$23, 67198.txt$\rightarrow$19, 73545.txt$\rightarrow$23, 74816.txt$\rightarrow$16)                                                                 \\ \hline
    Matthew Parker      & (64363.txt$\rightarrow$11)                                                   \\ \hline
    Daniel Frohman       & (63704.txt$\rightarrow$5, 66992.txt$\rightarrow$25, 69668.txt$\rightarrow$4, 68743.txt$\rightarrow$5, 67554.txt$\rightarrow$25, 67450.txt$\rightarrow$5, 72274.txt$\rightarrow$24, 69444.txt$\rightarrow$4)                                \\ \hline
   \end{tabular}  
\caption{Snapshot of People Gazetteer with Person names, Document list of occurrence and their corresponding Topic ID. The table shows Person entity recognized by PNER on the left along with a list of text articles in which s/he occurs on the right with the topic labels obtained for each of the text article during Topic Detection. The text articles in our corpus are indexed from 61102.txt to 74150.txt while the 30 topics are indexed from 0 to 29. The list of these topics can be viewed in Appendix Table ~\ref{table:topicwords}.}
\label{table:gaz}
\end{center}
  \end{minipage}
\end{table*}

%% file: IPDetection.tex
To find \emph{influential} people from the text in news articles, we define a score corresponding to each person entity in the gazetteer. This score is called the \emph{Influential Person Index (IPI)}. To calculate IPI, we first define a \emph{Document Index (DI)} to measure how each document in the person's associated list of documents affects his influence score. The choice of features for detecting whether a person is influential is motivated by following questions: (a) Are people mentioned frequently in the newspaper \emph{influential}? (b) How to measure frequency of occurrence(s) of a person - article by article or across the complete dataset? (c) Do longer documents talk more about important persons? (d) Is a person discussed in varied contexts (and over multiple topics in news) more influential than one who is consistently talked about in articles belonging to a single topic?  The following discussion describes the features chosen for calculation of DI and IPI of a person, followed by the complete algorithm for detection of influential persons.

\noindent \textbf{Document Index (DI): }
The Document Index (DI) of an article in the people gazetteer helps to measure a person's influence score. This is calculated for all the articles that a person entity is associated with in the People Gazetteer as follows:

\noindent \textbf{Normalized Document Length (NDL): } 
Document Length is defined as the number of tokens contained in a news article. It is further normalized by dividing with the maximum length of any news article across the corpus. Thus,  
$NDL= \frac{\text{Document Length}} {\text{Maximum Document Length in the dataset}}$


\noindent \textbf{ Normalized Person Name Frequency (NPNF): }
Person Name Frequency (PNF) accounts for the number of occurrences of a person's name in the news article. A high value of PNF makes the document more important. It is normalized as follows:

\begin{center}
$NPNF=	1+\log$(PNF)
\end{center}

The important questions that arise when dealing with Person Name Frequency are: (a) \textbf{Coreference resolution} of person names: (for e.g., names ``William Schmittberger",``Captain Williams" in a text article are same but recognized as separate persons) and (b) \textbf{Named Entity Disambiguation:} This refers to the occurrence of different persons with similar name in news articles. For e.g., the person ``John Smith" detected in two different articles may or may not refer to the same person. These issues are not dealt with by the PNER and need to be addressed separately. 
The following section explains the approach to coreference resolution.

\noindent \textbf{Coreference Resolution: } The coreference resolution aims to find all expressions that refer to the same person in the text. 
The algorithm used in this work uses a multi-pass sieve for coreference resolution\cite{lee2013deterministic} and consists of three steps: (a) Mention Detection: The goal is to  detect nouns, pronouns and occurrences of named entities from the text (b) Coreference Resolution: This is performed by using a combination of ten independent sieves applied from highest to lowest precision with global information sharing so that each sieve builds on the previously clustered mentions followed by post processing which removes singleton mentions. An implementation of this is available from the Deterministic Coreference Resolution System of the CoreNLP toolkit\footnote{http://nlp.stanford.edu/software/dcoref.shtml}. During resolution, all mentions that refer to the same entity are clustered together to form a coreference chain (c) Identification of the most representative mention: The most representative mention from each coreference chain is found and checked to ensure it is a valid person entity in the People Gazetteer. If it is found, then the count of mentions in its coreference chain is considered in the person name frequency determination. 

Table~\ref{table:coref}  illustrates an example of coreference resolution when applied to an article text from our corpus and its effect on PNF for a person entity from the People Gazetteer. 
 It is also observed that due to lack of punctuation in the dataset, several meaningless coreference mentions are also detected which are not person named entities. However, since the persons entities list is available from the People Gazetteer, frequencies of only those person names are replaced through Coreference Resolution. 


\begin{table*}
\centering
\begin{tabular} {|p{16cm}|}
\hline
\textbf{Text}: Mr Eugene Kelly bead of the banking house of Eugene Kelly A Co Is dying at his home
33 West Fiftyfirst street He became ill on Dee 4 and was forced to lake to his bed
\\ \hline

\textbf{Mentions Extracted}:
[``Mr Eugene Kelly bead of the banking house of Eugene Kelly A Co''], [``Eugene Kelly''],[``the banking house of Eugene Kelly A Co''], [``Eugene Kelly A Co''], [``Eugene Kelly''],[``his home 33 West Fiftyfirst street He became ill on Dee 4 and was forced to lake to his bed''], [``his''], [``33''], [``He''], [``Dee 4''], [``4''], [``his bed''], [``his'']
\\ \hline

\textbf{Coreferred Entity Chains with most representative entity in bold}:

\textbf{[``Mr Eugene Kelly bead of the banking house of Eugene Kelly A Co"]},

[\textbf{``Eugene Kelly"}, ``Eugene Kelly”, ``his", ``He", ``his"],

\textbf{[``the banking house of Eugene Kelly A Co"]},

\textbf{[``Eugene Kelly A Co"]},

\textbf{[``his home 33 West Fiftyfirst street He became ill on Dee 4 and was forced to lake to his bed"]},

\textbf{[``33"]},

\textbf{[``Dee 4"]},

\textbf{[``4"]},

\textbf{[``his bed"]}
\\ \hline
\textbf{Count of coreference mentions for most representative entity which is also a Person Named Entity}:

Eugene Kelly: 5 		\\ \hline
\end{tabular}
\caption{Table illustrating change in PNF for person named entities on using coreference resolution on an article text. The text has a person named entity ``Eugene Kelly'' with name frequency=2. During coreference resolution, all mentions are extracted from the text first and coreference chains are obtained. Most representative entity from each coreference chain is then identified (which may or may not be a person named entity) and if it matches any person named entity from the People Gazetteer, then the PNF for this entity is replaced by the count of mentions found in its coreference chain. Here, out of 9 coreference chains, ``Eugene Kelly'' is the most representative mention in one of the chains with 5 coreferences and since ``Eugene Kelly'' is also a person entity as verified from the People Gazetteer, we replace its PNF from 2 to 5.}
\label{table:coref}
\end{table*}

\noindent \textbf{Number of similar articles (NSIM): }
This parameter is used in the calculation of the DI by finding articles belonging to the same topic. 
For a document $d$ whose DI is to be calculated, let  SIM= Number of articles with the same topic as $d$ in the document list of a person.
This measure is normalized by the total number of articles in the document list of the person. Formally,

$NSIM= \frac{\text{SIM}} {\text{Total number of articles in the person's document list}}$
NSIM is equivalent to the proportion of topic similar articles that document $d$ has.

The Document Index is a function of the above mentioned parameters and is calculated by using the following formula:
\begin{center}

			$DI = w_a . NDL + w_b . NSIM + w_c . NPNF $
\end{center}
where, $w_a$, $ w_b$ and $w_c$ are the weights for NDL, NSIM and NPNF respectively.
DI is a heuristic measure -- each of the parameters can be weighted as per dataset characteristics and user requirements. For example, a higher value to $w_a$ and lower to $w_b$ and $w_c$ indicates documents with longer lengths are considered more important for influencing a person's IPI. On the other hand, a higher value to $w_b$ and lower to $w_a$ and $w_c$ indicates a document with larger proportion of topic similar articles influences the person's IPI more suggesting assignment of high influence score to a person entity occurring repeatedly in a specific news topic.  

\subsection{Influential Person Index (IPI)}

Once DI is calculated, the IPI is estimated as follows:
		
\begin{center}
$IPI= \text{max } DI(d_1, d_2, ...,d_n)+ UniqT$
\end{center}

where, max DI $(d_1,d_2...d_n)$ is the maximum DI found from a person's list of n articles, $UniqT = \frac{\text{Number of Unique Article Topics in a person entity's document list}}{\text{Total Number of Topics in the corpus}}$. The parameter $UniqT$  is used to account for the fact that a single person entity can be talked about multiple news topics in the news articles and to include its effect on the person entity's influence score. 
To rank people, the IPI are sorted in decreasing order. 
  
\subsection{Algorithm for Detection of Influential People (ADIP)}

Algorithm~\ref{algorithm:3} depicts the steps for measuring influence and ranking of influential people from the gazetteer. It starts with calculation of the Document Index for each news article in a person's document list. The weights $w_a$,$w_b$,$w_c$ are taken as inputs and multiplied with parameters $\text{NDL, NPNF and NSIM}$  to get the final estimate for DI. The list of DI scores is then sorted to find the maximum value amongst all news articles in the person's document list. The maximum DI score is then added to the UniqT parameter to get the final IPI for each person entity. Sorting the IPIs results in a ranked list of influential person entities.

\begin{algorithm*}[!th]
\begin{algorithmic}
\Function {CalculateIPI}{}

 \KwIn{$PeopleGazetter(Person Name,(DocList,TopicList))$, $w_a$,$w_b$,$w_c$}
\KwResult{Ranked List of Person Names}  
 $NPNF \leftarrow $0
 $NDL \leftarrow $0
 $NSIM \leftarrow $0
 $DI\leftarrow $0
 $UniqT\leftarrow $0
 $IPI\leftarrow $0\;  
  
    \For{(String PersonName : Persons)}
     {
	   \For{(String doc  : DocList)}
	{	
		$NPNF=1+\log (GetPNF(doc))$;
		
$NDL=GetDocLength(doc)/GetMaxDocLength()$;

		$ NSIM=GetTopicSimilarArticles(doc,DocList)$;

		$DI=w_a . NDL+w_b . NSIM+ w_c . NPNF$;
		
		$DIScoreList.add(DI)$;
 	 }
		$Sort(DIScoreList)$;

		$UniqT=GetUniqueTopics(Person,TopicList)$;

		$IPI=Max(DIScoreList)+UniqT$;

		$IPIScores.put(PersonName,IPI)$;
       }
	$Sort(IPIScores)$;

	$PrintPersonNameandIPI(IPIScores)$;

\EndFunction
\end{algorithmic}
\caption{Algorithm for Detection of Influential People (ADIP)}
\label{algorithm:3}
\end{algorithm*}

\begin{table*}
\begin{center}
\begin{tabular}{|c|c|} \hline
Function Name & Description \\ \hline
GetPNF(doc) & Calculates name frequency of the person entity \\
 & in document $doc$ \\ \hline
GetDocLength(doc) & Calculates number of tokens in $doc$. \\ \hline
GetMaxDocLength() & Calculates maximum number of \\
& tokens in any document.\\ \hline
GetTopicSimilarArticles(doc,DocList) &  Calculates normalized number \\
& of topic similar articles for $doc$ in the $DocList$. \\ \hline
Sort(DIScoreList) & Sorts the $DIScoreList$ \\ \hline
Max(DIScoreList) & Finds the maximum score from $DIScoreList$. \\ \hline
GetUniqueTopics(Person,TopicList) & Calculates normalized unique \\
& topics for $Person$ in its $TopicList$. \\ \hline
Sort(IPIScores) & Sorts the $IPIScores$ by IPI values. \\ \hline
PrintPersonNameandIPI(IPIScores) & Prints $Person$ name with its \\
& IPI in decreasing order of IPI value. \\ \hline
\end{tabular}
\caption{Description of the functions used in Algorithm~\ref{algorithm:3}}
\label{default}
\end{center}
\vspace{-10pt}
\end{table*}%

\begin{figure}
\begin{center}
\includegraphics[scale=0.75]{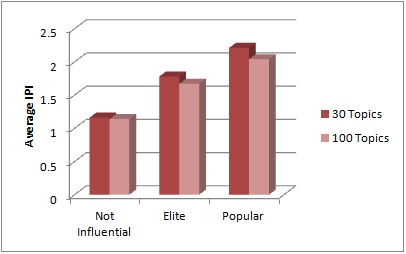}
\end{center}
\caption{Comparison of the Average IPI for two ranked lists $L_1$ and $L_2$ using $30$ and $100$ topics respectively.}
\label{figure:IPI}
\vspace{-10pt}
\end{figure}

%% file: Results.tex
\begin{table}[t!]
\resizebox{8cm}{!} {
\centering

    \begin{tabular}{|lllll|}
    \hline
    \textbf{Category}  &  \textbf{No. of }   & \textbf{Avg. No. of }   &  \textbf{Avg. Doc. }	&  \textbf{Avg. Person }	\\  
    & \textbf{People} & \textbf{Documents} & \textbf{Length} &  \textbf{Name } \\
    & & & & \textbf{Frequency} \\  \hline
Not Influ. & 36004 & 1.04 & 2119.6 & 1.07 	\\ \hline
 Popular & 344 & 5.75 & 1976.3 & 6.68  \\ \hline
Elite & 16 & 22.8 & 2971.5 & 29.870	 \\	\hline 
  \end{tabular}}
\caption {Table illustrating average statistics for each Person Category of the People Gazetteer.}
\label{table:stats}  
\end{table}

We present statistics pertaining to each category (not influential, popular and elite) of the people gazetteer in Table~\ref{table:stats}. 
It is observed that, on average the \emph{elite} are discussed very often in the news articles.
Document Length need not be high for a person to be more influential -- average document length obtained for the popular group is high in spite of their average IPI being low. This indicates that the number of similar articles for each document as well as the person name frequency play an important part in measuring influence.\\
\noindent \textbf{Effect of varying the number of topics in the topic model: } Our objective is to study the sensitivity of the ranked lists to parameters of the ADIP algorithm. We study the effect of the number of topics used in the LDA Model on the influential people list.
Two ranked lists of influential people, $L_1$ and $L_2$ are compared by using 30 and 100 topics\footnote{Several topic models are built by varying parameters of AD-LDA algorithm including number of iterations, topics and processors and their \emph{perplexity} is measured. Models with less perplexity are used for the study.} respectively. The weights $w_a$, $w_b$ and $w_c$ are set to 1 to ensure that all parameters have equal importance during calculation of DI and IPI. Figure~\ref{figure:IPI} shows the average IPI from the two ranked lists -- it appears that the average IPI for highly influential people is more susceptible to changes in number of topics.

 
\begin{table*}
\centering

\resizebox{15cm}{!}{
\begin{tabular}{|p{2cm}|l|p{1.5cm}|p{1.5cm}|l|l|l|p{3cm}|l|l|}
\hline
Person Name    & IPI  & Number of Articles & Person Category & NDL  & NPNF  & NSIM & TOPIC WORDS                                                          & UniqT & Rank \\ \hline
capt creeten   & 3.32 & 10                 & Popular          & 0.55 & 1.9 & 0.8  & mr court police judge justice case yesterday street district         & 0.06  & 1    \\ \hline
capt hankey    & 3.05 & 5                  & Popular          & 0.68 & 1.69  & 0.6 & club game team play football half ball left college back             & 0.06  & 2    \\ \hline
capt pinckney  & 2.93 & 3                  & Not Inf.        & 0.38 & 1.84 & 0.67 & man ho men night back wa room left house told bad                    & 0.03  & 3    \\ \hline
john martin    & 2.89 & 14                 & Popular          & 0.55 & 1.6  & 0.57  & mr court police judge justice case yesterday street district witness & 0.16  & 4    \\ \hline
ann arbor   & 2.87 & 44		    & Elite		& 0.19   & 1.77     & 0.63    &  dr book st story books cloth author cure free work york blood   illustrated remedy goods medical library health price & 0.26 & 5	\\  \hline
john macdonald & 2.85 & 3                  & Not Inf.        & 0.55 & 2.2  & 0    & great people life man women good country world american part         & 0.1   & 6    \\ \hline
aaron trow     & 2.81 & 1                  & Not Inf.        & 0.7  & 2.07 & 0    & man ho men night back wa room left house told                        & 0.03  & 6    \\ \hline
mrs oakes      & 2.79 & 5                  & Popular          & 0.08 & 2.04 & 0.6  & street mrs mr avenue wife house miss yesterday years home            & 0.06  & 7    \\ \hline
alexander iii  & 2.71 & 31                 & Elite            & 0.24 & 2.04 & 0.25 & great people life man women good country world american part         & 0.16  & 9    \\ \hline
buenos ayres   & 2.7 & 6                  & Popular          & 0.49 & 1.47  & 0.67 & white water indian black long found thu big dog time                 & 0.06  & 10    \\ \hline
\end{tabular}}
\caption{Table showing top 10 influential persons of List L1 detected from People Gazetteer with 30 Topics LDA model. Parameters NDL, NPNF, NSIM and Topic Words belong to the maximum scoring DI in the person's document list.}
\label{table:30t}

\end{table*}

\begin{table*}
\centering
\resizebox{15cm}{!} {
\begin{tabular}{|p{2cm}|l|p{1.5cm}|p{1.5cm}|l|l|l|p{3cm}|l|l|}
\hline
Person Name    & IPI  & Number of Articles & Person Category        & NDL  & NPNF  & NSIM & Topic Words                                                                 & UniqT & Rank \\ \hline
capt creeten   & 3.28 & 10                 & Popular     & 0.55 & 1.90 & 0.8  & mr police witness committee capt asked captain money inspector paid         & 0.06  & 1    \\ \hline
mrs martin     & 3.21 & 8                  & Popular     & 0.20 & 2.36 & 0.62  & mrs mr years wife home house ago woman city died                            & 0.02  & 2    \\ \hline
capt hankey    & 2.97 & 6                  & Popular    & 0.68 & 1.69  & 0.8 & game team football play half line ball back yale eleven                     & 0.02  & 3   \\ \hline
alexander iii  & 3.05 & 31                 & Elite     & 0.49 & 2.04 & 0.45 & emperor prince french alexander czar london nov government imperial russian & 0.07  & 4    \\ \hline
aaron trow     & 2.79 & 1                  & Not Inf. & 0.70 & 2.07 & 0    & day place long great water time feet found good men                         & 0.01  & 5    \\ \hline
john martin    & 2.78 & 14                 & Popular     & 0.55 & 1.6  & 0.57  & mr police witness committee capt asked captain money inspector paid         & 0.05  & 6    \\ \hline
john macdonald & 2.77 & 3                  & Not Inf. & 0.55 & 2.2  & 0    & people american man great country men world life good english               & 0.02  & 7    \\ \hline
mrs oakes      & 2.74 & 5                  & Popular     & 0.08 & 2.04 & 0.6  & mrs mr years wife home house ago woman city died                            & 0.02  & 7    \\ \hline
ed kearney     & 2.63 & 7                  & Popular     & 0.16 & 1.6  & 0.85 & won time race ran mile furlough half lo track fourth                        & 0.01  & 9    \\ \hline
caleb morton   & 2.61 & 1                  & Not Inf. & 0.70 & 1.9  & 0    & day place long great water time feet found good men                         & 0.01  & 10   \\ \hline

\end{tabular}}
\caption{Table showing top 10 influential persons of  List L2 detected from People Gazetteer with 100 Topics LDA model. Parameters NDL, NPNF, NSIM and Topic Words belong to the maximum scoring DI in the person's document list.}
\label{table:100t}
\vspace*{-10pt}
\end{table*}


The top 10 influential people from lists $L_1$ and $L_2$ are presented in Tables~\ref{table:30t} and ~\ref{table:100t} respectively. 
Our results suggest that none of the measures of NDL, NPNF or NSIM can be used alone to say whether a person is influential since these value do not decrease or increase consistently although the NPNF measure does contribute most to the IPI of any person.

The ranked lists $L_1$ and $L_2$ can be compared in terms of NSIM, UniqT and topic words to see the effect of 30 and 100 topics LDA models on influential person detection.
 If NSIM remains same in $L_1$ and $L_2$ during influential person detection, then the same highest scoring article DI is selected for calculation of IPI in both of them. This is why the parameters NDL and NPNF remain same across both the lists. This can be seen for ``capt creeten", ``capt hankey", ``aaron trow" and ``mrs oakes" in Tables~\ref{table:30t} and ~\ref{table:100t}. However, the value of UniqT for these persons decreases leading to decrease in their final IPI. This is because if the LDA model with higher number of topics (100) is used, the proportion of unique topics becomes lower when NSIM does not change. When the NSIM value changes because of change in number of topics, a different article with maximum DI score can get selected leading to change in the values of NDL, NPNF, UniqT and the final IPI. This causes a shift in the ranking of influential persons across the two lists and can be seen when the rank of ``alexander iii" in the first table moves from 9 to 4 in the second table. 
This helps to illustrate how a change in number of topics affects the ranking of influential people.

Wilcoxon signed rank paired test is also performed on the ranks of influential people across the two lists $L_1$ and $L_2$. This is done to test the hypothesis whether the differences in the ranking of person entities obtained using the 30 topic and 100 topic LDA models are due to chance or not. The null hypothesis for the test is: 
$H_0$: the distribution of difference of ranks of the persons across $L_1$ and $L_2$ is symmetric about zero. On performing the normal distribution approximation for 36364 samples of person ranks from lists $L_1$ and $L_2$, the results are found to be significant for both one-tail and two-tail tests.

\noindent \textbf{Case Studies: } To evaluate whether the ranking algorithm indeed finds people of \emph{influence}, manual evaluation of results is necessary. A description of the influential people from lists $L_1$ and $L_2$  (Table ~\ref{table:30t} and ~\ref{table:100t}) are discussed below: 
(1) \textbf{Elite} - This category as defined earlier includes people with greater than 16 news articles. However, only one person  (``alexander iii") from this category occurs in the top 10 influential persons. The entry for ``alexander iii" has an IPI of 2.71 and 3.05 respectively in lists $L_1$ and $L_2$ . The person occurs in 31 news articles with 5 and 7 different topics in each of the lists. The most common topic words associated with this person entity are: ``emperor prince french alexander czar london nov government imperial russian" indicating the importance of this entity in government related news topics. 
It is also observed that ``ann arbor" occurring in 44 articles is ranked 5 in list $L_1$ is a false positive as it is actually a location and has been wrongly ecognized in the PNER process as a person entity. 
(2)\textbf{Popular} - The top 10 influential entities from Tables~\ref{table:30t} and ~\ref{table:100t} contain the most number of people from this person category. The person entity ``capt creeten" has been ranked as highest influential (Rank 1) across both the tables. It occurs in 10 news articles with 9 of them belonging to the same topic indicating the person influencing news articles of high topic similarity. Some of the most common topic words for this entity include ``mr police witness committee capt asked captain money inspector paid" indicating the importance of this entity in a judicial or police related news topic.
Several persons from this category like ``mrs martin" , ``mrs oakes"  although identified among the top 10 influential persons suffer from the problem of named entity disambiguation as it is hard to identify which exact person they refer to due to lack of first names.
It is also observed that ``buenos ayres" occurring in 6 articles is ranked 10 in list $L_1$ is a false positive as it is actually a location and has been wrongly recognized in the PNER process as a person entity. 
 (3)\textbf{Not Influential} - Person entities belonging to this category have extremely low occurrence in news articles although the IPI of topmost influential entities belonging to this category are comparable to those in the other 2 categories.
Several person entities occurring in low number of news articles like ``aaron trow", ``caleb morton", ``john macdonald"  belong to this category. These entities in spite of occurring in very few articles (1 to 3) have high term frequency in those articles with comparatively longer article length indicating the importance of these entities with respect to the articles they occur in. Since each of the features has been given equal weight during the calculation of IPI,  these person entities with high NDL and NPNF have been identified among the top 10 influential persons. 

\subsection{Evaluation}

Due to the unavailability of ground truth consisting of influential people in the newspaper archives from November-December 1894, there is no way to validate our results. 
To broadly evaluate our results, a simple web search query with the person's name in the context of 19th century was done on the Wikipedia website for the top 30 influential persons of Lists $L_1$ and $L_2$ detected from the people gazetteer with 30 topics LDA and 100 topics LDA Model respectively.

\begin{figure*}
\begin{center}
\includegraphics[scale=0.75]{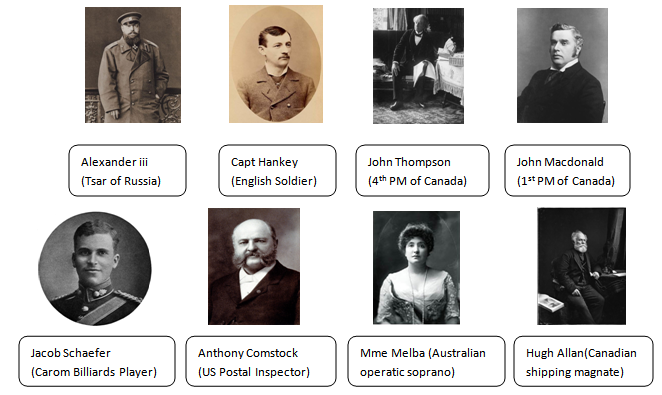}
\caption{Some of the top 30 influential persons obtained from the dataset and also found on Wikipedia during evaluation}
\label{figure:inf}
\vspace*{-10pt}

\end{center}
\end{figure*}

Among the top 30, 17 people from List $L_1$ and 12 from List $L_2$ were found to be influential and popular in the 19th century across topic categories like theatre, politics, government, shipping, etc. Some of these influential people have been found in Wikipedia and are shown in Figure~\ref{figure:inf}. Most of the false positives, although influential in other respects, were not  influential \emph{person} entities. This can be attributed to the incorrect PNER (Person Named Entity Recognition) on noisy OCR data. The ranked list of the top 30 influential persons with their IPI from Lists $L_1$ and $L_2$ can be seen in the Appendix (Tables ~\ref{table:app1} and ~\ref{table:app2}) where evaluation result for each person entity is also presented.

%% file: Discussion.tex
The following issues encountered during algorithm design and empirical verification are worth discussion:
(1)  A linear combination of parameters was used in our experiments for calculation of Document Index and Influential Person Index. Furthermore, they are weighted equally in experiments performed with our data. The heuristics can be re-weighted according to user requirements. 
(2) The parameters for calculation of DI and IPI can also be learned by performing regression analysis using a manually developed sample of influential people and obtaining the complete list of ranked influential people based on the learned parameters.
(3) The Normalized Document Length (NDL) defined for calculation of DI is normalized using the maximum length of any document in the dataset. However, there might exist other ways of normalization of Document Length like using total number of tokens in a person's document list or total number of tokens in the complete dataset which can be tuned based on the requirements of the application.
(4) Lists of influential people contain several false positives. This is due to noise in the OCR data -- several location and organization names have been recognized as person entities even after performing spelling correction resulting in false detection during PNER for some influential entities like ``van cortlandt", ``ann arbor" ,  and ``sandy hook". We recognize that there there is no gold standard data to measure the accuracy in our case therefore, we rely on the NER software's accuracy for recognizing person names.
(5) We have used the topic with highest probability for topic detection in the People Gazetteer. However, there could be other topics of interest. The algorithm can be modified to assign top N (N can be chosen depending on the topic distribution probabilities in the corpus) topics to a document by updating NSIM as follows: 
 $NSIM=  \frac{\sum_{i=1}^{N}\text{SIM}_i} { N\times n}$
Here, $N$=set of topics assigned to a document during topic detection,  $SIM_i$ =Number of articles with topic $i$ in the person's document list  and $n$= total number of documents in which person entity occurs in the People Gazetteer.
(6) The choice of parameters for topic detection also affects the detection of influential people which is evident from the fact that we get different ranking of influential people for the two different LDA topic model settings used.

\noindent \textbf {An alternative approach to detection of influential people: }
A heuristics based approach for finding influential people has been discussed in the previous section. It can be easily seen that the gazetteer comprises of a list of person names and a \emph{bag} of articles for each person from which his/her influence score can be learnt. This motivates discussion of whether an alternative approach involving multiple instance clustering (such as the \textbf{BA}g Level \textbf{M}ulti-\textbf{I}nstance \textbf{C}lustering (BAMIC \cite{zhang2009multi}) can be used for the problem. The procedure works as follows: 
 Cluster person entities into \emph{influential} or \emph{non-influential} categories by considering each person entity as having a bag of news articles in which their names occur.
The parameters used for calculating DI in the previous section i.e. NDL,NTF and NSIM can be used as features associated with each article instance in the bag. The group of \emph{influential} people identified can then be ranked using an appropriate ranking score.
The open source version of the BAMIC algorithm was used to compare results with the heuristic based approach proposed in this paper. However, the clustering algorithm does not scale very well. Our dataset consisted of roughly 40000 person named entities -- it was estimated that it would take around 200 days to get the clusters of influential and non-influential people.  Hence these results of comparison are not presented. 

%% file: Conclusion.tex
The problem of finding influential people from a historical OCR news repository has been studied to aid quantitative prosopography research. The solution framework comprises of development of a people gazetteer for facilitating the process of influential person detection. A novel algorithm for detecting influence has been presented which examines spelling correction of noisy OCR, person name entity recognition, topic detection and heuristics to measure influence and rank of people mentioned in the articles. Our algorithm has been tested on approximately 40000 people discussed in historic newspapers. The Tsar of Russia (Alexander III), the first and fourth Prime Ministers of Canada (John McDonald and John Thompson), and English soldiers serving in World War I (Captain Hankey) are among the influential people identified by the algorithm.

\section*{Acknowledgements}
This work was initially supported by the National Endowment of Humanities grant no. NEH HD-51153- 10. The authors would like to thank Barbara Taranto and Ben Vershbow from the NYPL Labs and Manoj Pooleery, Deepak Sankargouda and Megha Gupta for setting up the database used in this research.